\documentclass[preprint,nofootinbib,epsf]{revtex4}
\usepackage{amssymb}
\usepackage{graphicx}

\renewcommand{\thefootnote}{\fnsymbol{footnote}}
\newcommand{\vp}{\varepsilon}

\begin{document}
\preprint{BA-05-05}
\title{TeV Scale Leptogenesis, $\theta_{13}$ And Doubly Charged Particles At LHC}
\vskip 8cm
\author{Shahida Dar}
\author{Qaisar Shafi}
\author{Arunansu Sil}
\affiliation{Bartol Research Institute, Department of Physics and Astronomy, 
University of Delaware, Newark, DE 19716, USA}
\begin{abstract}
We explore a realistic supersymmetric $SU(2)_L\times SU(2)_R\times U(1)_{B-L}$ model spontaneously
broken at around $10^{12}$ GeV. The presence of $D$ and $F$-flat directions
gives rise to TeV mass doubly charged particles which can be found at the LHC.
We implement TeV scale leptogenesis and employing both type I and II
seesaw, the three light neutrinos are partially
degenerate with masses in the $0.02-0.1$ eV range. The effective mass parameter for
neutrinoless double beta decay is $0.03-0.05$ eV. We also find the interesting 
relation $\tan 2 \theta_{13}\simeq \frac{\sin 2 \theta_{12}}{\tan 2\theta_{23}}
\left(\frac{\Delta m^2_{\odot}}{\Delta m^2_{\rm{atm}}}\right)\,\lesssim 0.02$. 
\end{abstract}
\maketitle  

\renewcommand{\thefootnote}{\arabic{footnote}}
\setcounter{footnote}{0}

It has been recognized for some time that spontaneously broken supersymmetric
models with $D$ and $F$-flat directions can lead to interesting phenomenological
and cosmological consequences \cite{Lazarides:1985bj, Lazarides:1985ja, 
Yamamoto:1985rd, Binetruy:ss, Lazarides:1986rt, Lazarides:1987yq, 
Lazarides:1992gg}. An intermediate symmetry breaking scale
of order $10^{8}-10^{16}$ GeV is a characteristic feature of these models. Another
important aspect is the appearance of thermal inflation typically involving
about 10 or so {\it e}-foldings \cite{Lazarides:1985ja, Binetruy:ss, 
Lazarides:1987yq, Lazarides:1992gg}. 
The entropy generation associated with thermal inflation has been exploited to try to resolve the 
gravitino \cite{Lazarides:1985ja} and moduli problem \cite{Lyth:1995ka}, 
and to suppress the primordial monopole number density to
acceptable levels \cite{Lazarides:1986rt}. 
The entropy production does have an important drawback
though. It will dilute , and in some cases completely wash away, any pre-
existing baryon asymmetry. This very depends on the magnitude of the
intermediate scale $M_{I}$ \cite{Lazarides:1987yq}.

In \cite{Dar:2003cr}, with an intermediate scale of order $10^{8}$ GeV, the observed baryon
asymmetry was explained via resonant leptogenesis \cite{Pilaftsis:2003gt}. The relatively low
intermediate scale causes a moderate amount of dilution of an initially large
lepton asymmetry, such that the final baryon asymmetry is consistent with the
observations. This paper is partly motivated by the desire to implement
TeV scale leptogenesis in models with an intermediate scale that is 
higher, namely of order $10^{12}$ GeV. (Scales significantly higher than 
this lead to a reheat temperature after thermal inflation that is too low for 
sphaleron transitions to be effective). To be specific, we base our
discussion on the supersymmetric version of the well known gauge group
$G_{221} = SU(2)_{L} \times SU(2)_{R} \times U(1)_{B-L}$ \cite{Pati:1974yy, Mohapatra: 1974hk}. 
The presence of $D$ and $F$-flat
directions means that the `flaton' fields $\phi$, $\overline {\phi}$ with vevs $= M_{I} \sim 10^{12}$
GeV, have an associated mass scale $M_{s}$, the supersymmetry breaking scale, of
the order of TeV. Being in the triplet representation of $SU(2)_{R}$, these fields
contain doubly charged particles which turn out to have masses of order $M_{s}$. Hence, 
they should be found at the LHC. 

Thermal inflation is driven by $\phi, \overline{\phi}$ and after it
is over, the flatons produce TeV mass right-handed neutrinos associated 
with the first two families, whose subsequent decay leads via 
leptogenesis to the observed baryon asymmetry. (Because it has mass
of order $M_{I}$, the third generation right-handed neutrino is not accessible at
the TeV scale). Taking into account both type I \cite{Mohapatra:1979ia} and type II \cite{Lazarides:1980nt} 
seesaw mechanism, the light neutrinos turn out to have partially degenerate \cite{king} masses 
close to $0.02-0.1$ eV (depending upon the choice of $M_s$ and $M_I$), 
consistent with solar and atmospheric neutrino mass scales and mixings. The third neutrino mixing 
angle is given by $\sin \theta_{13} \lesssim 0.01$, while the effective mass parameter associated 
with neutrinoless double beta decay is about $0.03-0.05$ eV.

The gauge symmetry $G_{221}$ is broken at an intermediate scale $M_I$ which arises 
from an interplay between the supersymmetry breaking scale $M_s \ll M_I$ and a cutoff scale 
$M_*$. Since $G_{221}$ is broken to the gauge group $SU(2)_{L} \times U(1)_{Y}$ with the vevs of 
$\phi\,(1, 3, -2)$ and $\overline {\phi}\,(1, 3, 2)$, a discrete $Z_{2}$ symmetry 
remains unbroken \cite{Kibble:1982ae} which is precisely `matter' parity. Consequently, the LSP 
is stable. To generate the scale $M_I$ via $D$ and $F$-flat directions, 
we employ a discrete symmetry $Z_4 \times Z_8$ which, among other things, 
prevents terms such as $\overline{\phi}\phi$ from appearing in the superpotential.   

Consider the superpotential
\begin{eqnarray}\label{wnr}
W &\supset& \kappa H H \frac{(\overline{\phi} \phi)^3}{M_{*}^{5}}\,+\,
\frac{\lambda}{4} \frac{(\overline{\phi} \phi)^4}{{M_{*}^5}}\,+
\,\gamma_{33} L_3^c L_3^c \phi\,+\,\gamma_{12} L_1^c L_2^c \phi 
\left(\frac{\overline{\phi}\phi}{M_{*}^{2}}\right)^2 \nonumber \\
&+&Y_{22} L_2 L_2^c H\,+\,Y_{33}L_3 L_3^c H\,+\,Y_{13} L_{1} 
L_{3}^c H \frac{(\overline{\phi}\phi)}{{M_{*}^2}}\,+
\,pL_1 L_1 \Delta_L\,+\,\frac{c}{M_*} \Delta_L \phi H H \nonumber \\
&+& \frac{a}{M_*}\overline{\Delta}_L \Delta_L \overline{\phi}\phi
\,+\, \frac{b}{M_*} \overline{\phi}\,\overline{\Delta}_L H H 
\left(\frac{\overline{\phi} \phi}{M_*^2}\right)^2
\,+\, \frac{d}{M_*} \overline{\Delta}_L\overline{\Delta}_L\overline{\phi}\,\overline{\phi}
\left(\frac{\overline{\phi} \phi}{M_*^2}\right)^2\,,
\end{eqnarray}
where $H$ ($\equiv [H_u, H_d]$) is a bidoublet\footnote{With only one bidoublet higgs, there will be no CKM 
mixings. However, $G_{221}$ can be embedded in a bigger group such as 
$SO(10)$, and it is then possible to induce non-zero CKM mixings through some additional `matter fields' 
\cite{Shafi:1998yy}. Another possibility is to include loop contributions in association with 
supersymmetry breaking terms \cite{rnckm}. To simplify our presentation we will not address these
possibilities here. Generation of the lepton-mixing matrix will be discussed later in the paper.} higgs 
superfield (2,2,0), $L_i$ (2,1,-1), $L_j^c$ (1,2,1) are the left-handed and 
right-handed lepton doublets ($i,j=1, 2, 3$ are family indices) respectively, 
$\Delta_L$ (3, 1, 2) is a $SU(2)_L$ higgs 
triplet with conjugate superfield $\overline{\Delta}_L$ (3, 1, -2) which will be needed later 
on to provide type II seesaw contribution to the light neutrino mass matrix. 
The term proportional to $\kappa$ can help resolve the MSSM $\mu$ problem 
($\mu \sim \kappa \left(\frac{M_I}{M_*}\right)^5 M_I$ (see also \cite{Dar:2003cr}), 
and is expected to be of order few hundred GeV). With $\kappa$ and $\gamma_{12}$ of comparable magnitudes, 
the dominant decay channel of $\phi$ is to first and second generation right-handed neutrinos, 
$N_1, N_2$, and this is useful in realizing TeV scale leptogenesis through $N_{1,2}$ decay. 
In Table I we list the discrete charges of the various superfields. (For the cosmology of spontaneously 
broken discrete symmetries the reader is referred to \cite{Lazarides:1992gg}.)
\begin{table}[b]
\begin{center}
\begin{tabular}{|c||c||c||c||c||c||c||c||c||c||c||c|}
\hline
Fields & $\phi$ & $\overline{\phi}$ & $L_1^c$ & $L_2^c$ & $L_3^c$ & 
$H$ & $L_1$ & $L_2$ & $L_3$ & $\Delta_L$ & $\overline{\Delta}_L$\\
\hline
$Z_4$ & 1 & -1 & 1 & 1 & 1 & $i$ & $i$ & $-i$ & $-i$ & $-1$ & $1$\\
$Z_8$ & $\omega^4$ & $\omega^6$ & $\omega$ & $\omega^7$ & $\omega^2$ & $\omega$ & $\omega^3$ & 1 & $\omega^5$ 
& $\omega^2$ & $\omega^4$\\
\hline
\end{tabular}
\end{center}
\caption{\small Discrete charges of various superfields.}
\end{table}

The zero temperature effective scalar 
potential of $\phi$ (we use $\phi$ to also represent the scalar 
component of the superfield) along the $D$-flat direction 
with $\langle \phi 
\rangle=\langle\overline{\phi}\rangle ^{\dag}$, is given by
\begin{equation}
V(\phi)\,=\,\mu_{0}^{4}-2M_{s}^{2}\left|\phi\right|^2+2 \frac{\lambda^{2}{\left|\phi\right|^{14}}}{M_{*}^{10}}\,,
\end{equation}
where $\mu_{0}^{4}$ is introduced to ensure that at the 
minimum $\langle\phi\rangle=M_I$, $V(M_I)=0$, and $-2M_s^2 \left|\phi\right|^2$ 
is the soft supersymmetry breaking term with $M_s \sim$ TeV. Here it is assumed
that a positive supersymmetry breaking mass squared term generated at some 
superheavy scale can acquire a negative sign, via radiative corrections involving 
the superpotential coupling $\gamma_{33}L_3^c L_3^c \phi$, at a lower energy \cite{Lazarides:2000em}.
Minimization of the effective potential yields the intermediate scale,
\begin{equation}\label{im}
M_{I}=\left|\langle\phi\rangle\right|=\left[\frac{M_{s}^{2}M_{*}^{10}}
{7 \lambda^{2}}\right]^{1/12}. 
\end{equation}
We will see shortly that $M_I \simeq 10^{12}$ GeV and $M_* \simeq 5.5 \times 10^{13}$ GeV, for 
$M_s \simeq 5.5$ TeV are compatible with TeV scale lepton asymmetry, with partial conversion of 
the latter via sphalerons into the observed baryon asymmetry.

For non-zero temperature ($T$) the effective potential gets an additional 
contribution \cite{Dolan:1973qd}, $V_{T}(\phi)$ \cite{Dar:2003cr}. 
For $\phi\ll T$ the temperature-dependent mass term is $\sigma \,T^{2}\big|\phi\big|^{2}$, 
where $\sigma \simeq 0.14$. For $T > T_c = \sqrt{2/\sigma} M_s$ the potential
\begin{equation}\label{vt}
V(\phi)\,=\,\mu^4_0+(-2 M_{s}^{2}+\sigma T^{2})
\big|\phi\big|^{2}+2\frac{ \lambda^{2}}{M_{*}^{10}}\,\big|\phi\big|^{14}\,,
\end{equation}
develops a minimum at $\phi=0$, with  $V(\phi=0)\,=\,\mu^4_0 = \frac{12}{7} M_s^2 M_I^2$.  
For $\phi > T$, the temperature-dependent term is 
exponentially suppressed and $V(\phi)$ develops another minimum 
at $\phi = M_I$. $\phi =0$ remains an absolute minimum for $\mu_0 \lesssim 
T \lesssim M_I$, but for $T \lesssim \mu_0$, the true minimum at $M_I$ takes over. 
The dark energy density associated with the absolute minimum 
($10^{-12}$ eV$^4$) is irrelevant for our purpose \cite{Shafi:2005fs}.  

Due to the false vacuum energy density $\mu_{0}^4$ the universe 
experiences roughly $\ln(\mu_0/T_c)\sim 8$ {\it e}-foldings 
of thermal inf\mbox{}lation. The flaton has mass $m_{\phi}$ of order $2 \sqrt {6} M_s$ 
and it can decay into right-handed neutrinos (with mass $M_N$) via the superpotential 
coupling $\gamma_{12} L_1^c L_2^c \phi \left(\overline{\phi} \phi/M_{*}^{2}\right)^2$. 
The decay width is given by 
\begin{equation}\label{gphi}
\Gamma_{\phi}\simeq\frac{\gamma_{12}^2}{8 \pi}\,
m_{\phi}\,\left(\frac{M_I}{M_*}\right)^8 f_{\phi}\,=\,0.04\,\gamma_{12}^2\,
M_s\,\left(\frac{M_s}{\lambda M_I}\right)^{8/5} f_{\phi}\,,
\end{equation}
where we have used Eq. (\ref{im}) with $m_{\phi}$ in terms of $M_s$, 
and $f_{\phi}=(1-4M_N^2/m_{\phi}^2)^{3/2}$. The other decay 
width, $\Gamma_{\phi \rightarrow {\tilde H} \tilde H} \sim (\kappa^2/8 \pi) (M_I/M_*)^{10} m_{\phi}$, is 
clearly suppressed compared to $\Gamma_{\phi}$. 
Therefore the final temperature ($T_f$) after thermal inflation can be expressed as 
$T_f\simeq 0.3 \sqrt{(\Gamma_{\phi} M_{P})} \sim 0.06\,\gamma_{12} (M_s/\lambda M_I)^{4/5} 
\sqrt{(M_{s} f_{\phi} M_{P})}$, where $M_{P}=2.4 \times 10^{18}$ GeV is the reduced Planck mass. 
To estimate $T_f$ we need to know $\gamma_{12}$ which can be estimated as follows. 

From the superpotential in Eq. (\ref{wnr}) the right-handed neutrino mass matrix $M_R$ 
is given by 
\begin{equation}\label{mr}
M_R=\left( \begin{array} {ccc}
0 & x & 0\\
x & 0 & 0\\
0 & 0 & M\\
\end{array} \right), 
\end{equation}
where $x= \gamma_{12} \left(M_{I}/M_{*}\right)^4 M_I= \gamma_{12} M_I^{1/5}M_s^{4/5} 
\left(1/\sqrt{7} \lambda\right)^{4/5}$ and $M=\gamma_{33} M_I$. Diagonalizing $M_R$, 
we have $M_R = U^*_{RR} M^{diag}_R U^{\dag}_{RR}$, where 
\begin{equation}\label{ur}
U_{RR}\,=\,\left( \begin{array}{ccc}
1/\sqrt{2} & 1/\sqrt{2} & 0\\
-1/\sqrt{2} & 1/\sqrt{2} & 0\\
0 & 0 & 1\\
\end{array} \right).~P\,,
\end{equation}
with real and positive eigenvalues $\left|M_{1}\right|=\left|M_2\right| = \left|x\right|$ 
and $\left|M_3\right| = \left|M\right|$. The phases of $M_i$ are absorbed
in $U_{RR}$ through the phase matrix $P = diag (e^{-i\alpha_{1}/2}, 
e^{-i\alpha_{2}/2}, e^{-i\alpha_{3}/2})$ with $\alpha_{i}=$ arg $\left(M_i\right)$. 
Obviously the decay of flaton to $N_{1,2}$ is possible if $m_{\phi} \gtrsim 2 M_{1,2}$ i. e. 
$\left|\gamma_{12}\right| \lesssim \sqrt{6} (\sqrt{7} \lambda)^{4/5}
\left(\frac{M_s}{M_I}\right)^{1/5}\,$.

Hence in this model we find the final temperature to be  
\begin{equation}\label{tf}
T_f\,\simeq\,90\,{\rm {GeV}}\left(\frac{\gamma_{12}}{0.074}\right)\,
\left(\frac{M_s}{5.5\,{\rm {TeV}}}\right)^{13/10}\,
\left(\frac{10^{12}\rm {GeV}}{M_I}\right)^{4/5}\,\left(\frac{1}{\lambda}\right)^{4/5}\,,
\end{equation}
with $M_N/m_{\phi} \simeq 0.3$. It is gratifying that $T_f$ is in a range
where the electroweak sphalerons \cite{Fukugita:1986hr} are able to convert some 
fraction of the lepton asymmetry into baryon asymmetry \cite{Kuzmin:1985mm}, which sets 
an upper bound on $M_I$ of $10^{12}$ GeV for $M_s \sim$ few TeV. 
In Fig. (1), the variation of $T_f$ with $M_I$ (for fixed $M_s$ and $M_{1,2} \simeq 0.3\,m_{\phi}$) 
is shown where Eq. (\ref {gphi}) is used along with the constraint on $|\gamma_{12}|$. 
For example with $|\gamma_{12}| \sim 0.074$ (for $\kappa \sim |\gamma_{12}|$, the 
$\mu$ term is ${\cal O}(10^2)$ GeV), the masses of first two 
generations of right-handed neutrinos are of order 8 TeV corresponding to $T_f \simeq 90$ GeV. 

We now consider the case where $N_{1, 2}$ are produced by the direct non-thermal decay of the 
flaton field $\phi$. The ratio of the number density of 
right-handed neutrino, $n_N$, to the entropy density $s$ is given by
\begin{equation}
\frac{n_N}{s}\simeq \frac{3}{2} \frac{T_f}{m_{\phi}} B_r\,,
\end{equation}
where $B_r$ denotes the the branching ratio into the right-handed neutrino channel. 
The resulting total lepton asymmetry produced by the $N_1, N_2$ decay is $\frac{n_L}{s}
=\sum_{i} \frac{n_N}{s} \epsilon_i$, where $\epsilon_i$ is the lepton asymmetry produced 
per $i$th right-handed neutrino decay. 

Unlike thermal leptogenesis, there is no wash-out factor in this non-thermal scenario \cite{Lazarides:1991wu} 
corresponding to the lepton number violating 2-body scatterings mediated by right-handed neutrinos, 
as long as the light right-handed neutrino masses $|M_i| \gg T_f$ 
\cite{Fukugita:1990gb, Fukugita:1992xc}. The wash-out factor is proportional to $e^{-z}$, where 
$z=M_i/T_f$ \cite{Buchmuller:2003gz} and for $z \gtrsim 10$ it can be safely neglected. 

The CP asymmetry $\epsilon_i$ is given by \cite{Luty:1992un}
\begin{equation}\label{ei}
\epsilon_i =\frac{1}{8 \pi} \sum_{k\neq i} f\left( \frac{\left|M_{k}\right|^2}{\left|M_i\right|^2}
\right)\, \frac{{\rm{Im}} \left[(h^{\dagger}h)_{ik}^2\right]}{(h^{\dagger}h)_{ii}}\,,
\end{equation}
where 
\begin{equation}\label{fy}
f(y)\,=\,\sqrt{y} \left[\frac{2}{1-y}\,-\,{\rm{ln}} \left(1+\frac{1}{y}\right)\right]\,,
\end{equation}
and $h=\frac{m_D'}{v}$ is the neutrino Yukawa coupling matrix in the basis where the right-handed 
neutrino mass matrix $M_R$ is diagonal with real and positive eigenvalues and 
$v$ is the electroweak scale vev $\left(\simeq 174 \, {\rm GeV} \right)$. This expression is valid in the limit 
where $\left|\left|M_i\right| - \left|M_j\right|\right| > \Gamma_{N_i} + \Gamma_{N_j} $, 
where $\Gamma_{N_i}=\frac{1}{8 \pi}(h^{\dagger}h)_{ii}\left|M_i\right|$ 
represents the decay width of the $i$th right-handed neutrino, and applies in our case. 

The mass degeneracy between $\left|M_1\right|$ and $\left|M_2\right|$ can be broken by assuming the existence
of new physics beyond $M_* (\simeq 5.5\times 10^{13}$ GeV$)$ which does not respect the discrete symmetry. Assuming
this scale, $M_G$, to be near the GUT scale\footnote{One may wonder about the origin of the 
two scales $M_*$ and $M_G$. In a scenario with extra dimension(s) they can be associated with the 
compactification and cutoff scales respectively.}, an additional term in the
superpotential such as $W_{\xi}=\eta L_2^c L_2^c \phi (\overline\phi \phi/M^2_{G})$
can provide a suitable splitting ($\xi=\eta M_I (M_I/M_G)^2$), which can lead to the desired lepton asymmetry. To 
estimate the latter, let us assume that in the basis with $M_R$ given by Eq. (\ref{mr}), the 
Dirac mass matrix $m_D$ can be diagonalized by a bi-unitary transformation   
\begin{equation}\label{mdd}
m_D\, =\, U_L^{\dagger} m_D ^{diag} U_{RD}\, ,
\end{equation}
which leads to 
\begin{equation}\label{hh}
h^{\dagger}h\,=\,\frac{1}{v^2} U_R^{\dagger} \left(m_D^{diag}\right)^2 U_R\,, 
\end{equation}
where $U_R=U_{RD}U_{RR}$. We will consider $m_D^{diag}\equiv diag(m_{D_1}, m_{D_2}, 
m_{D_3})=diag (m_e, m_{\mu}, m_{\tau}) \tan\beta $ which is possible within a 
left-right framework \cite{Lazarides:1993sn, Lazarides:1993uw}. The diagonal entries are taken to be real and positive. 
Notice that the left-handed rotation $U_L$ is not present in Eq. (\ref{hh}). From Eq. (\ref{wnr}), 
$m_D$ is given by 
\begin{equation}\label{md}
m_D=\left( \begin{array} {ccc}
0 & 0 & \vp_1\\
0 & Y_{22}v & 0\\
0 & 0 & Y_{33}v\\
\end{array} \right)\,, 
\end{equation}
where $\vp_1=Y_{13}v\left(M_I/M_*\right)^2$. We find that 
$U_{RD} = I$, along with $m_{D_1}=0$, $m_{D_2}\simeq Y_{22} v=m_{\mu}\tan \beta$ 
and $m_{D_3}\simeq Y_{33} v = m_{\tau}\tan \beta$ as previously mentioned. (Note that the electron mass 
vanishes in this approximation. It could arise via radiative corrections through 
flavor violating supersymmetry breaking contributions \cite{Ferrandis:2004ng}.) The deviation of $U_L$ 
from identity matrix is parameterized by a small angle $\theta_{13_L}$ proportional to $\vp_1/m_{D_3} \lesssim 10^{-2}$. 

Substituting $U_{RR}$ from Eq. (\ref{ur}) into Eq. (\ref{hh}), we find  
\begin{equation}\label{hdh}
\left(h^{\dagger}h\right)_{12}\,=\,\frac{
-m_{D_{2}}^{2}}{2 v^2}\, e^{i(\alpha_{1}-\alpha_{2})/2}\,;~~~\left(h^{\dagger}h\right)_{11}\,=
\,\left(h^{\dagger}h\right)_{22}\,=\,\frac{m_{D_{2}}^{2}}{2 v^2}\,.
\end{equation}
In the limit of degenerate neutrinos $y\rightarrow 1$ and thus $f(y) \simeq 2/(1-y)$. From Eq. (\ref{ei}) we then have
\begin{eqnarray}\label{ep}
\epsilon_1 \simeq \epsilon_2\,&\simeq&\,\frac{1}{8 \pi}\,\frac{\left|M_1\right|}{\left|M_1\right|-\left|M_2\right|}\, 
\frac{{\rm{Im}} \left[\left(h^{\dagger}h\right)_{12}^2\right]}{\left(h^{\dagger}h\right)_{11}}\,, \nonumber\\
&\simeq& \frac{\Gamma_{N_1}}{\left|M_{1}\right|}\,\,\frac{\left|M_1\right|}{\left|M_1\right|-
\left|M_2\right|}\,\,\sin(\alpha_1-\alpha_2)\,, 
\end{eqnarray}
where we have used Eq. (\ref{hdh}). Recall that the mass degeneracy between 
$\left|M_1\right|, \left|M_2\right|$ is removed by the introduction of the $\xi$ term\footnote{The introduction of new contributions 
like $W_{\xi}$ and additional terms in $m_D$ 
(to be considered later) which are important for light neutrino mixings, will not have much 
impact on Eqs. (\ref{hdh}) and (\ref{ep}), and the results for lepton asymmetry will remain unaffected.} with the 
superpotential $W_{\xi}$. 

Note that the enhancement of the lepton asymmetry through the near degeneracy 
of $\left|M_1\right|$ and $\left|M_2\right|$ is restricted not only by the condition 
\begin{equation}\label{mg}
\left||M_1|-|M_2|\right| \gtrsim 2 \Gamma_{N_1}\,,
\end{equation}
but also by their nearly opposite CP parities. In the limit $|\xi| \ll \left|x\right|$, 
so that $\left|M_{1(2)}\right|\simeq \left|x\right| \left[1-(+) \frac{|\xi|}{|x|}
\cos (\delta_x-\delta_{\xi})+\frac{1}{2}\frac{|\xi|^2}{|x|^2} \right]$, 
Eq. (\ref{ep}) can be translated\footnote{In this limit, sin$
(\alpha_1-\alpha_2)\simeq \frac{|\xi|}{|x|}\frac{\sin(\delta_x-\delta_{\xi})}{\left[1-
\frac{|\xi|^2}{|x|^2}\cos^2(\delta_x-\delta_{\xi})\right]}$.} into \cite{Akhmedov:2003dg}
\begin{equation}
\epsilon_1\simeq\epsilon_2\simeq \frac{\Gamma_{N_1}}{2 \left|M_1\right|}\, \tan(\delta_{x}-\delta_{\xi})\,,
\end{equation}
where $\delta_x-\delta_{\xi}=\frac{1}{2}$arg$(\frac{x^2}{\xi^2})$. To achieve $|\xi|\ll |x|$, we need to consider 
$\eta \lesssim 10^{-2}$. Using equality condition \cite{Pilaftsis:1998pd} from Eq. (\ref{mg}), 
one can estimate the maximum value of $\epsilon=\epsilon_1+\epsilon_2$ to be 
$\frac{\left|\xi\right|}{\left|M_1\right|}$sin$(\delta_x-\delta_{\xi}) \simeq 
\frac{\left|\xi\right|}{\left|M_1\right|}$. The lepton asymmetry, with $B_r \sim 1$, is then 
\begin{equation}\label{fnl}
\frac{n_L}{s}\simeq \frac{3}{2}\,\frac{T_f}{m_{\phi}}\cdot 2\cdot \epsilon_1\,\simeq\,\frac{3}{2}
\, \frac{T_f}{m_{\phi}}\,\frac{\Gamma_{N_1}}{\left|M_1\right|}\, \tan(\delta_x-\delta_{\xi})\,.
\end{equation}
From the observed baryon to photon ratio $n_B/n_{\gamma} \simeq (6.5 \pm  0.4) \times 10^{-10}$ \cite{Spergel:2003cb} 
the lepton asymmetry is found to be $|n_L/s| \simeq (2.67-3.02) \times 10^{-10}$, where we have used $n_B/s \simeq 
(n_B/n_{\gamma})/7.04$ \cite{Kolb:1990vq} and $n_L/s = - (111/36) n_B/s$ \cite{Khlebnikov:1988sr}
\footnote{Here, the final temperature is just below the electroweak crossover scale, and we follow 
\cite{Khlebnikov:1988sr} to estimate the approximate conversion factor relating 
lepton and baryon asymmetries as 36/111 $\simeq 0.324$. We thank the referee for raising this point.}. 

Before discussing the magnitudes of the parameters involved in Eq. (\ref {fnl}) in order to be consistent with 
the observed $n_B/s$, let us first consider the light neutrino 
masses and related issues. From solar, atmospheric and terrestrial neutrino data (at 95\% C.L.) 
\cite{Nakaya:2002ki}, we have  
\begin{equation}\label{data}
\begin{array}{l}
\Delta m^2_{\odot}\equiv\Delta m^2_{12} = 7.92\, (1\pm 0.09)\times 10^{-5}\mathrm{\ eV}^2\;,~~~~
\sin^2\theta_{12} = 0.314 \,(1^{+0.18}_{-0.15})\;;\\ 
\Delta m^2_{atm}\equiv\Delta m^2_{23}= 2.4\,(1^{+0.21}_{-0.26})
\times 10^{-3}\mathrm{\ eV}^2\;,~~~~
\sin^2\theta_{23} =0.44\,(1^{+0.41}_{-0.22})\;; \\
\sin^2\theta_{13} = 0.9^{+2.3}_{-0.9}\times 10^{-2}\;.
\end{array}
\end{equation}

Our next task is to make sure that the light neutrino mass 
matrix $m_{\nu}$ is consistent with Eq. (\ref{data}). 
The lepton mixing matrix \cite{pmns} is given by $U_{PMNS} = U_L U_{\nu}$, where $U_L$ 
arises from the charged lepton sector, and $U_{\nu}$ comes from the diagonalization of $m_{\nu}$, namely 
$m^{diag}_{\nu} = U^{T}_{\nu} m_{\nu} U_{\nu}$. Since $\theta_{13_L} \lesssim 10^{-2}$, the bilarge 
mixings must arise from $U_{\nu}$. The structure of $m_D$ given in Eq. (\ref{md}) must be 
modified to generate appropriate atmospheric and solar neutrino mixings. Consider the terms $Y_{11} L_1 L_1^c H 
(\overline{\phi} \phi/M_G^2)$ and $Y_{31} L_3 L_1^c H (\overline{\phi} \phi/M_G^2)$ which contribute 
$\vp_3 \equiv Y_{11} v \left(M_I/M_G\right)^2 $ and $\vp_2 \equiv Y_{31}v \left(M_I/M_G\right)^2$ to the 
11 and 31 elements of $m_D$ respectively. Being sufficiently small, they leave intact the relations 
$m_{D_{2}}\simeq Y_{22} v = m_{\mu}\tan\beta$, $m_{D_{3}}\simeq Y_{33} v =m_{\tau}\tan\beta$, 
$m_{D_{1}}\sim 0$ (to leading order). The deviation of $U_{RD}$ from the identity 
matrix is parameterized by $\theta_{{13}_R} \sim \vp_2/m_{D_3}\ll 1$.

Including the type II seesaw contribution to the neutrino mass matrix from the induced vev of $\Delta_L$, 
we have\footnote{The scalar triplets in $\Delta_L$ are heavy with mass $\sim a M_I (M_I/M_*)$. 
Their decay does not contribute to the lepton asymmetry 
which requires two pairs of such triplets \cite{Ma:1998dx}.} 
\begin{eqnarray}
m_{\nu} &\simeq& a_{\Delta}\,-\,m_D M_R^{-1} m_D^T \,,\nonumber \\
&\simeq & \left( \begin{array} {ccc}
a_{\Delta}+\frac{\xi}{x^2}\vp_3^2 ~~~&~~~ -\frac{m_{D_{2}}}{x}\vp_3 ~~~&~~~ \frac{\xi}
{x^2}\vp_2\vp_3-\frac{m_{D_{3}}}{M_I}\vp_1\\
-\frac{m_{D_{2}}}{x}\vp_3 ~~~&~~~ 0 ~~~&~~~ -\frac{m_{D_{2}}}{x}\vp_2\\
\frac{\xi}{x^2}\vp_2\vp_3-\frac{m_{D_{3}}}{M_I}\vp_1 ~~~&~~~ -\frac{m_{D_{2}}}{x}\vp_2 ~~~&~~~ \frac{\xi}
{x^2}\vp_2^2-\frac{m_{D_3}^2}{M_I}\\
\end{array}\right)\,,
\end{eqnarray}
where, for simplicity,  $a_{\Delta}=2\,p\,\langle \Delta_L \rangle \simeq 2\,p\,(cd/|a|^2) 
(v^2/M_I) (M_I/M_*)^4$ is taken to be real. Note that the terms proportional to $\xi$ 
(with $\eta \sim 10^{-2}\,, |\xi| \sim 100$ GeV) are accompanied by factors 
$\vp_i\vp_j/x^2,~i, j=2, 3$, and can be safely ignored. 
As for the lepton asymmetry, only the relative phase between $\xi$ and $x$ will be important.

To obtain the mass eigenstates, we first rotate $m_{\nu}$ by $U'=U_{23}U_{13}$ 
(where $U_{ij}$ denotes the rotation matrix in the $ij$ sector and we will ignore CP violation here) and 
express the effective mass matrix 
${\widetilde m}_{\nu}= U'^{T} m_{\nu} U'$ in the new basis as
\begin{equation}\label{mnu}
{\widetilde m}_{\nu}\simeq \left( \begin{array} {ccc}
a_{\Delta}c_{13}^2+2 \rho_1 s_{13}c_{13}+\Lambda_{+} s_{13}^2 ~~~~&~~~~ \rho_2 c_{13} & 0\\ 
\rho_2 c_{13}~~~~ &~~~~ \Lambda_{-}~~~~ &~~~~\rho_2 s_{13}\\
0 ~~~~&~~~~ \rho_2 s_{13}~~~~&~~~~ a_{\Delta}s_{13}^2-2 \rho_1 s_{13}c_{13}+\Lambda_{+} c_{13}^2 \\
\end{array}\right)\,,
\end{equation}
where 
\begin{eqnarray}\label{del}
\Lambda_{+(-)}&=&-\,\frac{m_{D_3}^2}{M_I}c_{23}^2(s_{23}^2)-(+) 2 \frac{m_{D_{2}}}{x}\vp_2c_{23}s_{23}\,,\nonumber \\
\rho_{1(2)}&=& \frac{m_{D_{3}}}{M_I}\vp_1c_{23}(s_{23})+(-)\frac{m_{D_{2}}}{x}\vp_3s_{23}(c_{23})\,,
\end{eqnarray}
and $c_{ij}=\cos \theta_{ij}$ and $s_{ij}=\sin \theta_{ij}$. Furthermore, 
\begin{eqnarray}\label{t23}
\tan2\theta_{23}&=& \frac{2 m_{D_{2}} \vp_2}{x}\,\frac{M_I}{m_{D_3}^2}\,, \\
\tan2\theta_{13}&=&\frac{2\rho_1}{a_{\Delta}-\Lambda_{+}}\, \label{t13}.
\end{eqnarray}

Note that for $2 m_{D_2}\vp_2/x \gg m_{D_3}^2/M_I$, 23 mixing can be maximized. 
An approximate diagonalization of ${m}_{\nu}$ is achieved by focusing on the 
12 block of ${\widetilde m}_{\nu}$ and noting that $\theta_{13}$ is relatively small 
(see Eq. (\ref{data})). With $\rho_1 \ll ({a_{\Delta}-\Lambda_{+}})$, to a good approximation 
the third state in Eq. (\ref{mnu}) decouples. The upper left $2 \times 2$ block 
of ${\widetilde m}_{\nu}$ is readily diagonalized and the resulting mass eigenvalues are

\begin{equation}\label{mnu1}
m_{\nu_{1, 2}}\simeq \frac{1}{2}\left[\left(a_{\Delta}+\Lambda_{-}\right)\mp
\sqrt{\left(\Lambda_{-}-a_{\Delta}\right)^2+4 \rho_2^2}\right]\,;~~~m_{\nu_3}\simeq\Lambda_{+}\,,
\end{equation}
with 
\begin{equation}\label{t12}
\tan 2 \theta_{12}\simeq \frac{2 \rho_2}{\Lambda_{-}-a_{\Delta}}\,.
\end{equation}
Barring cancellation between $a_{\Delta}$ and $\Lambda_{-}$ a large but
non-maximal mixing angle $\theta_{12}$ is possible. The light neutrinos turn out to be partially degenerate $(|m_{\nu_1}| \sim |m_{\nu_2}| \sim |m_{\nu_3}| 
\gtrsim \sqrt{\Delta m^2_{\rm{atm}}})$. 
 
\begin{itemize}
\item The above consideration does not alter the requirement of $\tan \theta_{13}$ being small as 
$a_{\Delta} \sim \Lambda_{-} = - \Lambda_{+} 
\tan^2 \theta_{23}$. Here we have used compact forms of $\Lambda_+$ and $\Lambda_-$ 
with the help of Eq. (\ref{t23}): $\Lambda_+ =  - (m_{D_2}/x) \vp_2 \cot \theta_{23}; ~~ 
\Lambda_- = (m_{D_2}/x) \vp_2 \tan \theta_{23}$. 
Hence Eq. (\ref{t13}) can be approximated as 
\begin{equation}\label{simple13}
\tan 2 \theta_{13} \simeq \rho_1 \frac{x}{m_{D_2} \vp_2} \sin 2 \theta_{23}.
\end{equation}
\end{itemize}
\vskip 0.1 cm
\begin{table}[h]
\begin{center}
\begin{tabular}{|c|c|c|c|c|c|c|c|c|c|}
\hline
$\vp_1$ (GeV) & $\vp_2$ (GeV) &$\vp_3$ (GeV) & $a_{\Delta}$ (GeV) & $x$ (TeV)&$\tan \beta$ & $\tan 2\theta_{12}$ &
$\tan 2\theta_{23}$ & $\tan 2\theta_{13}$\\
\hline
\hline
$5.75 \times 10^{-3}$ & $1.67\times 10^{-6}$ & $1.74\times 10^{-8}$& $4.38 \times 10^{-11}$
 & 8.3 & 3.15 &  2.36 & 5.4 & $ 7 \times 10^{-3}$\\
\hline
\end{tabular}
\end{center}
\caption{\small A viable set of values for $\vp_i$ and the corresponding mixing angles 
(using $m_{\mu} \simeq 0.083$ GeV and $m_{\tau} \simeq 1.4$ GeV at $M_I$ \cite{Das:2000uk})}
\end{table}

Table II presents a set of parameters and the corresponding mixing angles for $M_I=10^{12}$ GeV 
and $M_s=5.5$ TeV. (To achieve $a_{\Delta} \sim 0.044$ eV, we take $|a|\sim 1.25 \times 10^{-2}$ 
and $c \sim d \sim p \sim {\cal O}(1)$).
The mass splittings are given by 
\begin{eqnarray}
\Delta m^2_{\odot} &\equiv& \Delta m_{12}^2=\left||m_{\nu_{1}}|^2-|m_{\nu_{2}}|^2\right|
\simeq (a_{\Delta}^2-\Lambda_{-}^2)\sqrt{1+\tan^{2}2\theta_{12}}\,,\nonumber \\
&\simeq& 4 a_{\Delta}\,\frac{|\rho_2|}{\sin 2 \theta_{12}}\,,
\end{eqnarray}
\begin{eqnarray}
\Delta m^2_{\rm{atm}} &\equiv& \Delta m_{23}^2 = \left||m_{\nu_{3}}|^2-|m_{\nu_{1}}|^2\right|\simeq 
\left|\Lambda_{+}^2-\Lambda_{-}^2-\rho_{2}^2+\frac{1}{2} \Delta m^2_{\odot}\right|\,, \nonumber \\
&\simeq& 2 a_{\Delta}\frac{m_{D_{2}}\vp_2}{x} \frac{\cot 2 \theta_{23}}{\sin^2 \theta_{23}}+
{\cal O}(\Delta m^2_{\odot})\,,
\end{eqnarray}
where we have used Eqs. (\ref{del}), (\ref{t23}), (\ref{mnu1}) and (\ref{t12}).

As the dominant contributions to $\rho_{1,2}$ come from the second term in their expressions, 
we have $|\rho_2| \simeq |\rho_1| \cot \theta_{23}$. Using Eq. (\ref{simple13}) we find 
\begin{equation}\label{relation}
\tan 2 \theta_{13}\simeq \frac{\sin 2 \theta_{12}}{\tan 2\theta_{23}}
\left(\frac{\Delta m^2_{\odot}}{\Delta m^2_{\rm{atm}}}\right)\,\lesssim 0.02\,.
\end{equation}
\begin{itemize}
\item The 13 mixing angle is well below the upper limit allowed by experiments. 
This is due to the fact that $\Delta m^2_{\odot}$ depends upon $|\rho_2|$ and in turn on 
$\tan 2 \theta_{13}$. Higher values of $\tan \theta_{13}$ near the experimental upper limit 
cannot reproduce the appropriate $\Delta m^2_{\odot}$. With the parameters in Table II, 
the mass-squared differences are $\Delta m^2_{\odot}\sim 7.6 \times 10^{-5}$ eV$^2$ 
and $\Delta m^2_{\rm{atm}} \sim 2 \times 10^{-3}$ eV$^2$. In Fig. (2) the allowed region 
for $\vp_2, \vp_3$ is shown for fixed $M_s$ and $M_I$. 
\item For simplicity we have taken $Y_{13}$ associated with $\vp_1$ to be $\lesssim 10^{-1}$ 
and the relation (\ref{relation}) holds to a good accuracy. We have checked numerically that 
$Y_{13}\sim {\cal O}(1)$ would not change anything except that somewhat higher values of $\vp_3$ 
($\sim 3.3 \times 10^{-8}$ GeV) are allowed.
\item Finally, $\theta_{13_L}$ induces a tiny correction to $\theta_{12}$ and $\theta_{23}$. 
$\theta_{13}$ receives a correction of order $\theta_{13_L} c_{23}$ which could be significant for
$Y_{13}$ of order unity, i.e. $\sin \theta_{13}^{\rm {eff}} \simeq \sin \theta_{13}- (\vp_1/m_{D_3}) c_{23}$. Even 
for this case the prediction for $\theta_{13}$ remains unaltered.
\end{itemize}

From Eq. (\ref{mnu1}) we find that the light neutrino masses are of the same order, close to $0.02-0.1$ eV. 
Figs. (3) and (4) display the range of allowed values for $m_{\nu_{1}}$ and $m_{\nu_{3}}$. Following 
\cite{king} these are partially degenerate neutrinos. This range of neutrino masses is below the so-called 
quasi-degenerate case. Furthermore, the effective mass parameter in neutrinoless double 
beta decay (which is the $ee$ element ($\equiv a_{\Delta}$) of the neutrino 
mass matrix) \cite{Klapdor-Kleingrothaus:2000sn} is estimated to be of order $0.03-0.05$ eV, corresponding 
to $M_I=10^{12}$ GeV and $M_s=5.5$ TeV.   

We have checked that renormalization effects \cite{Haba:1998fb} do not alter 
our conclusions in any significant way. The estimated splitting 
between $N_1$ and $N_2$ due to running from $M_*$ to $M_{1,2}$ 
is of order $\frac{m_{D_2}^2}{4 \pi^2 v^2}
{\rm {ln}}(\frac{M_*}{10^4 {\rm {GeV}}}) M_2$, 
which is much smaller than the contribution arising from the
term proportional to $\xi$. Furthermore, the running in $m_D$ can be
absorbed through a rescaling of $m_{D_i}$.

With the specified range of parameters involved, we are now in a position to calculate 
$n_L/s$ from Eq. (\ref{fnl}). Table III presents a sample value of the phase involved in $n_L/s$ which is 
required to produce correct amount of lepton asymmetry. All other parameters are taken from Table II.   

\begin{table}[h]
\begin{center}
\begin{tabular}{|c|c|c|}
\hline
$\gamma_{12}$ & $\tan(\delta_{x}-\delta_{\xi})$ &
$\frac{n_L}{s}$\\
\hline
\hline
0.074 &1.24 & $2.8 \times 10^{-10}$\\
\hline
\end{tabular}
\end{center}
\caption{\small Parameter values used  in order to produce the required lepton
asymmetry.}
\end{table}

An important feature of our model is the existence of TeV scale doubly charged 
particles \cite{Chacko:1997cm}. Writing 
\begin{equation}
\phi=\left[ \begin{array} {cc}
\frac{\phi^{-}}{\sqrt{2}} & \phi^0 \\
\phi^{--} & -\frac{\phi^{-}}{\sqrt{2}}\\
\end{array} \right]\,\,\,
\rm{and}\,\,\,\,
\overline{\phi}=\left[ \begin{array} {cc} 
\frac{\overline{\phi}^{+}}{\sqrt{2}} & \overline{\phi}^{++} \\
\overline{\phi}^{0} & -\frac{\overline{\phi}^{+}}{\sqrt{2}}\\
\end{array} \right], 
\end{equation}
and letting $\phi^0= M_I+ \eta/\sqrt{2}$ and $\overline{\phi}^0= M_I + \overline{\eta}/\sqrt{2}$ 
($\eta =\overline{\eta}^*$ is the real flaton field), 
the mass-squared matrix for the doubly charged particle is 
\begin{eqnarray}
\bordermatrix{ & {\phi}^{++} & {\overline {\phi}}^{++} \cr
{\phi}^{--} & \frac{1}{7} M_s^2+M_s^2
& \frac{6}{7} M_s^2 \cr
{\overline {\phi}}^{--} & \frac{6}{7} M_s^2 &  \frac{1}{7} M_s^2+M_s^2 \cr},
\end{eqnarray}
where Eq. (\ref {im}) has been used\footnote{The contributions from the $D$ 
terms turn out to be quite small.} and the soft masses are $M_s ^2 $ [Tr ($\phi^{\dagger} \phi$+
$\overline {\phi}^{\dagger}\overline{\phi}$)]. Hence, the lightest doubly charged particles have 
masses $\sim \sqrt{2/7}M_s \sim$ 2.9 TeV and can be found at the LHC \cite{Azuelos:2005uc}. The 
heavier doubly charged particles have masses of order 8 TeV. 
Note that the signs of the soft masses are important to keep 
the eigenvalues of this mass squared matrix positive. The existence of these light 
doubly charged states can also be inferred from the presence 
of an associated higher symmetry of the superpotential in Eq. (\ref{wnr}) which leads to pseudo-Goldstone 
bosons. Finally, we note that in addition to the full MSSM spectrum of fields, the 
model contains a new singly charged field of mass $\sqrt{2} M_s \simeq 8$ TeV. There is yet 
another singly charged field with mass of order $M_I$, well beyond the reach of LHC.

In summary, we have presented a realistic supersymmetric model with gauge symmetry 
$SU(2)_L \times SU(2)_R \times U(1)_{B-L}$, 
broken at an intermediate scale $M_I \sim 10^{12}$ GeV. Thermal inflation is followed by 
TeV scale leptogenesis. To reproduce the observed baryon asymmetry, the two lightest right-handed 
neutrinos are closely degenerate in mass, with $M_1 \simeq M_2 \sim 10^4$ GeV, 
while the mass of the third right-handed neutrino 
is $M_3 \sim 10^{12}$ GeV. The physics of neutrino oscillations requires both type I and type II 
seesaw, and the three light neutrinos turn out to be partially degenerate with 
masses around $0.02-0.1$ eV. This is close to the value of the mass parameter 
associated with neutrinoless double beta decay \cite{Klapdor-Kleingrothaus:2000sn}. An important test of the 
model is the presence of doubly charged particles that should be found at the LHC. 
Another important feature is the prediction $\sin \theta_{13} \lesssim $ 0.01. It would be of some interest to 
extend the discussion to larger gauge groups such as $SU(3)^3$ 
\cite{Lazarides:1993sn, Lazarides:1993uw, Gursey:1975ki} and $SU(4)_c \times 
SU(2)_L \times SU(2)_R$ \cite{Pati:1974yy}.

Q. S and A. S thank Rabi Mohapatra for fruitful discussion. 
A. S and S. D would like to thank M. Frigerio for useful correspondence. Q. S. also acknowledges 
the hospitality of the Alexander Von Humboldt Foundation and Professors 
H. Fritzsch, K. Koller and D. Luest. This work was supported by the U.S. DOE 
under Contract No. DE-FG02-91ER40626. 


\begin{figure}[t]
\begin{center}
\includegraphics[angle=270,width=13cm]{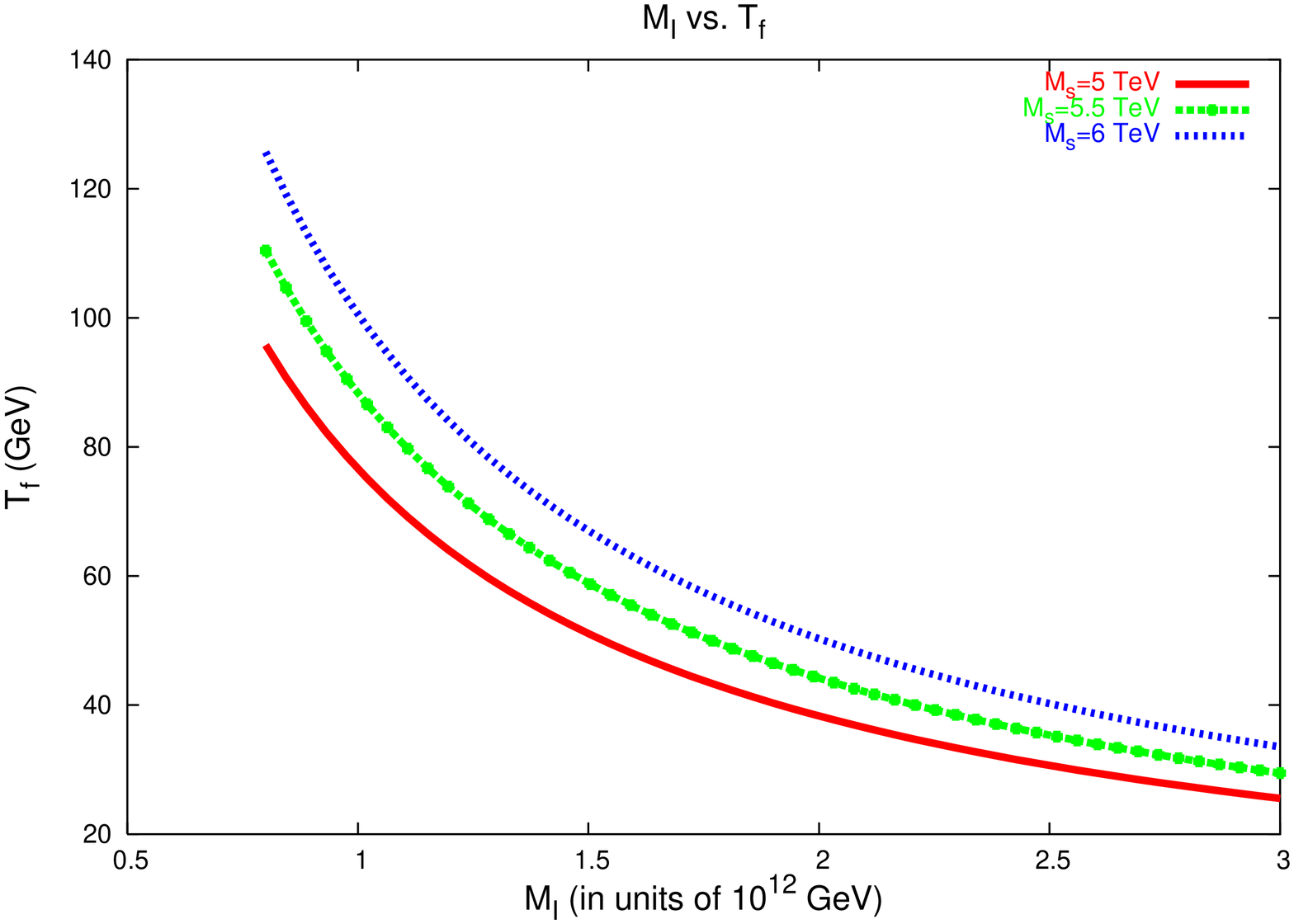}
\caption{Variation of $T_f$ with $M_I$ for different values of $M_s$.}
\end{center}\label{fig1}
\end{figure}

\begin{figure}[b]
\begin{center}
\includegraphics[angle=270,width=13cm]{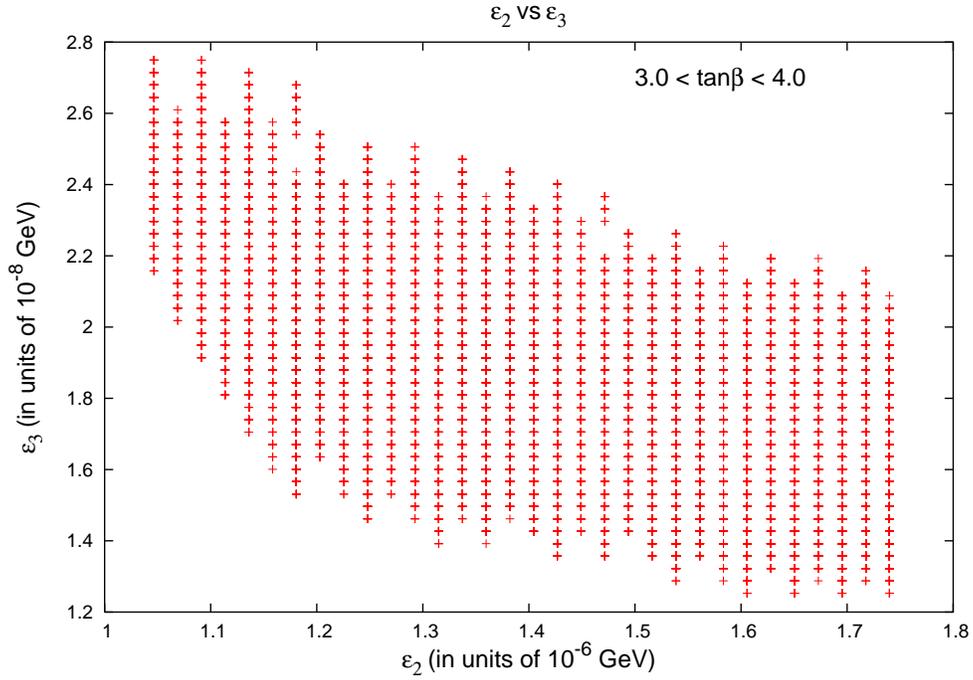}
\caption{The allowed region for $\vp_2$ and $\vp_3$
for $M_s=5.5$ TeV and $M_I=10^{12}$ GeV.}
\end{center}\label{fig2}
\end{figure}

\begin{figure}[t]
\begin{center}
\includegraphics[angle=270,width=13cm]{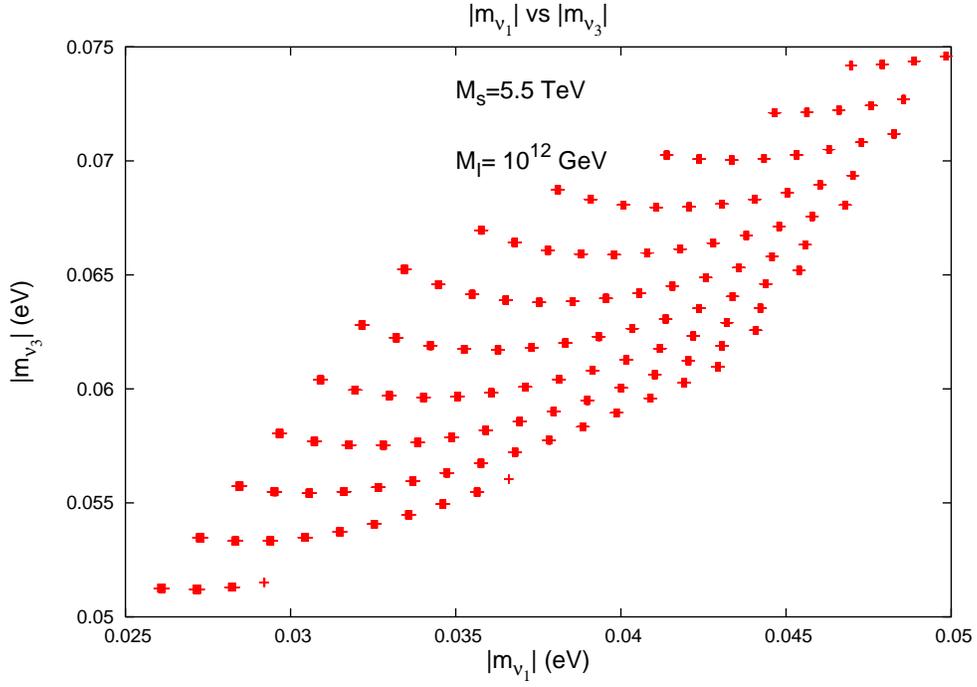}
\caption{The allowed region for $|m_{\nu_1}|$ and $|m_{\nu_3}|$.}
\end{center}\label{fig3}
\end{figure}

\begin{figure}[b]
\begin{center}
\includegraphics[angle=270,width=13cm]{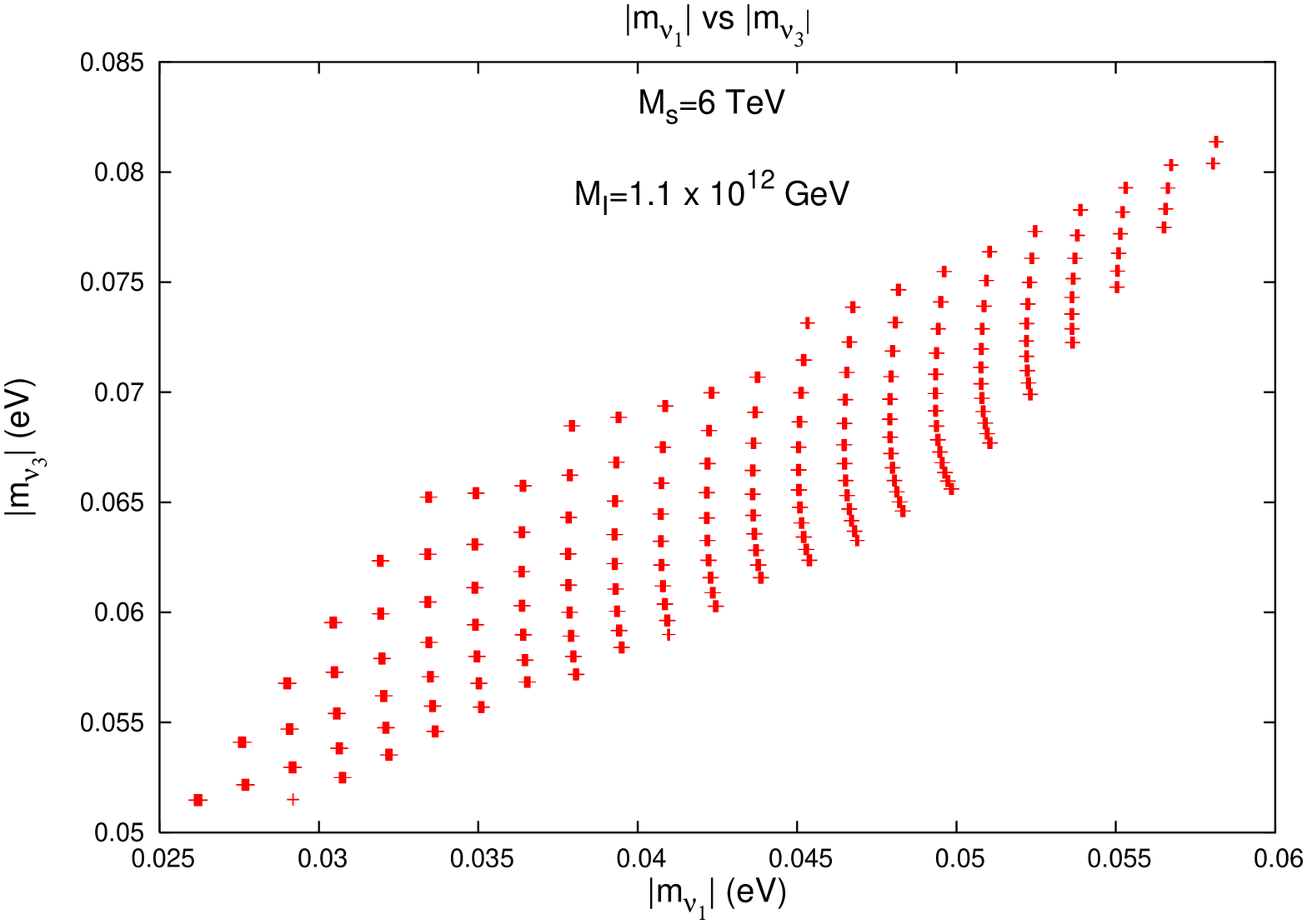}
\caption{Same as Fig. (3), but for different $M_s$ and $M_I$.}
\end{center}\label{fig4}
\end{figure}

\begin{thebibliography}{xx}

\bibitem{Lazarides:1985bj}
G.~Lazarides, C.~Panagiotakopoulos and Q.~Shafi,
Phys.\ Rev.\ Lett.\  {\bf 56}, 432 (1986).

\bibitem{Lazarides:1985ja}
G.~Lazarides, C.~Panagiotakopoulos and Q.~Shafi,
Phys.\ Rev.\ Lett.\  {\bf 56}, 557 (1986).

\bibitem{Yamamoto:1985rd}
K.~Yamamoto,
Phys.\ Lett.\ B {\bf 168}, 341 (1986).

\bibitem{Binetruy:ss}
P.~Binetruy and M.~K.~Gaillard,
Phys.\ Rev.\ D {\bf 34}, 3069 (1986).

\bibitem{Lazarides:1986rt}
G.~Lazarides, C.~Panagiotakopoulos and Q.~Shafi,
Phys.\ Rev.\ Lett.\  {\bf 58}, 1707 (1987).

\bibitem{Lazarides:1987yq}
G.~Lazarides, C.~Panagiotakopoulos and Q.~Shafi,
Nucl.\ Phys.\ B {\bf 307}, 937 (1988).

\bibitem{Lazarides:1992gg}
G.~Lazarides and Q.~Shafi,
Nucl.\ Phys.\ B {\bf 392}, 61 (1993).

\bibitem{Lyth:1995ka}
D.~H.~Lyth and E.~D.~Stewart,
Phys.\ Rev.\ D {\bf 53}, 1784 (1996)
[arXiv:hep-ph/9510204].

\bibitem{Dar:2003cr}
S.~Dar, S.~Huber, V.~N.~Senoguz and Q.~Shafi,
Phys.\ Rev.\ D {\bf 69}, 077701 (2004)
[arXiv:hep-ph/0311129].

\bibitem{Pilaftsis:2003gt}
A.~Pilaftsis and T.~E.~J.~Underwood,
Nucl.\ Phys.\ B {\bf 692}, 303 (2004)
[arXiv:hep-ph/0309342]; T.~Hambye, J.~March-Russell and S.~M.~West,
JHEP {\bf 0407}, 070 (2004)
[arXiv:hep-ph/0403183].

\bibitem{Pati:1974yy}
J.~C.~Pati and A.~Salam,
Phys.\ Rev.\ D {\bf 10}, 275 (1974).

\bibitem{Mohapatra: 1974hk}
R.~N.~Mohapatra and J.~C.~Pati,
Phys.\ Rev.\ D {\bf 11}, 566 (1975); G.~Senjanovic and R.~N.~Mohapatra,
Phys.\ Rev.\ D {\bf 12}, 1502 (1975); M.~Magg, Q.~Shafi and C.~Wetterich,
Phys.\ Lett.\ B {\bf 87}, 227 (1979); M.~Cvetic,
Nucl.\ Phys.\ B {\bf 233}, 387 (1984); C.~S.~Aulakh, A.~Melfo and G.~Senjanovic,
Phys.\ Rev.\ D {\bf 57}, 4174 (1998)
[arXiv:hep-ph/9707256]; G.~R.~Dvali, G.~Lazarides and Q.~Shafi,
Phys.\ Lett.\ B {\bf 424}, 259 (1998) [arXiv:hep-ph/9710314].

\bibitem{Mohapatra:1979ia}
P.~Minkowski,
Phys.\ Lett.\ B {\bf 67}, 421 (1977); T. Yanagida, in {\it Proceedings of the Workshop on Unified Theory and the 
Baryon Number of the Universe}, edited by O. Sawada and A. Sugamoto (KEK, 1979);
M. Gell-Mann, P. Ramond, and R. Slansky, in {\it Supergravity}, edited by F. van Nieuwenhuizen 
and D. Freedman (North Holland, Amsterdam, 1979); S.L. Glashow, Cargese Lectures (1979); 
S.~Weinberg,
Phys.\ Rev.\ Lett.\  {\bf 43}, 1566 (1979); R.~N.~Mohapatra and G.~Senjanovic,
Phys.\ Rev.\ Lett.\  {\bf 44}, 912 (1980).


\bibitem{Lazarides:1980nt}
G.~Lazarides, Q.~Shafi and C.~Wetterich,
Nucl.\ Phys.\ B {\bf 181}, 287 (1981); 
R.~N.~Mohapatra and G.~Senjanovic,
Phys.\ Rev.\ D {\bf 23}, 165 (1981); J.~Schechter and J.~W.~F.~Valle,
Phys.\ Rev.\ D {\bf 22}, 2227 (1980).

\bibitem{king}
S.~Antusch and S.~F.~King,
Nucl.\ Phys.\ B {\bf 705}, 239 (2005)
[arXiv:hep-ph/0402121].

\bibitem{Kibble:1982ae}
T.~W.~B.~Kibble, G.~Lazarides and Q.~Shafi,
Phys.\ Lett.\ B {\bf 113}, 237 (1982); S.~P.~Martin,
Phys.\ Rev.\ D {\bf 54}, 2340 (1996)
[arXiv:hep-ph/9602349].



\bibitem{Shafi:1998yy}
Q.~Shafi and Z.~Tavartkiladze,
Nucl.\ Phys.\ B {\bf 549}, 3 (1999)
[arXiv:hep-ph/9811282].

\bibitem{rnckm}
K.~S.~Babu, B.~Dutta and R.~N.~Mohapatra,
Phys.\ Rev.\ D {\bf 60}, 095004 (1999)
[arXiv:hep-ph/9812421].

\bibitem{Lazarides:2000em}
G.~Lazarides and Q.~Shafi,
Phys.\ Lett.\ B {\bf 489}, 194 (2000)
[arXiv:hep-ph/0006202].

\bibitem{Dolan:1973qd}
L.~Dolan and R.~Jackiw,
Phys.\ Rev.\ D {\bf 9}, 3320 (1974);
S.~Weinberg,
Phys.\ Rev.\ D {\bf 9}, 3357 (1974).

\bibitem{Shafi:2005fs}
Q.~Shafi, A.~Sil and S.~P.~Ng,
Phys.\ Lett.\ B {\bf 620}, 105 (2005)
[arXiv:hep-ph/0502254].
and references therein.

\bibitem{Fukugita:1986hr}
M.~Fukugita and T.~Yanagida,
Phys.\ Lett.\ B {\bf 174}, 45 (1986).

\bibitem{Kuzmin:1985mm}
V.~A.~Kuzmin, V.~A.~Rubakov and M.~E.~Shaposhnikov,
Phys.\ Lett.\ B {\bf 155}, 36 (1985).

\bibitem{Lazarides:1991wu}
G.~Lazarides and Q.~Shafi,
Phys.\ Lett.\ B {\bf 258}, 305 (1991).

\bibitem{Fukugita:1990gb}
M.~Fukugita and T.~Yanagida,
Phys.\ Rev.\ D {\bf 42}, 1285 (1990); D.~J.~H.~Chung, E.~W.~Kolb and A.~Riotto,
Phys.\ Rev.\ D {\bf 60}, 063504 (1999)
[arXiv:hep-ph/9809453].

\bibitem{Fukugita:1992xc}
M.~Fukugita, H.~Murayama, K.~Suehiro and T.~Yanagida,
Phys.\ Lett.\ B {\bf 283}, 142 (1992).

\bibitem{Buchmuller:2003gz}
W.~Buchmuller, P.~Di Bari and M.~Plumacher,
Nucl.\ Phys.\ B {\bf 665}, 445 (2003)
[arXiv:hep-ph/0302092].

\bibitem{Luty:1992un}
M.~A.~Luty,
Phys.\ Rev.\ D {\bf 45} (1992) 455; M.~Flanz, E.~A.~Paschos and U.~Sarkar,
Phys.\ Lett.\ B {\bf 345}, 248 (1995)
[Erratum-ibid.\ B {\bf 382}, 447 (1996)]
i[arXiv:hep-ph/9411366]; M.~Plumacher,
Z.\ Phys.\ C {\bf 74}, 549 (1997)
[arXiv:hep-ph/9604229]; L.~Covi, E.~Roulet and F.~Vissani,
Phys.\ Lett.\ B {\bf 384}, 169 (1996)
[arXiv:hep-ph/9605319]; W.~Buchmuller and M.~Plumacher,
Phys.\ Lett.\ B {\bf 431}, 354 (1998)
[arXiv:hep-ph/9710460].


\bibitem{Lazarides:1993sn}
G.~Lazarides, C.~Panagiotakopoulos and Q.~Shafi,
Phys.\ Lett.\ B {\bf 315}, 325 (1993) [Erratum-ibid.\ B 
{\bf 317}, 661 (1993)] [arXiv:hep-ph/9306332].

\bibitem{Lazarides:1993uw}
G.~Lazarides and C.~Panagiotakopoulos,
Phys.\ Lett.\ B {\bf 336}, 190 (1994)
[arXiv:hep-ph/9403317].


\bibitem{Ferrandis:2004ng}
J.~Ferrandis,
Phys.\ Rev.\ D {\bf 70}, 055002 (2004)
[arXiv:hep-ph/0404068].

\bibitem{Akhmedov:2003dg}
E.~K.~Akhmedov, M.~Frigerio and A.~Y.~Smirnov,
JHEP {\bf 0309}, 021 (2003)
[arXiv:hep-ph/0305322].

\bibitem{Pilaftsis:1998pd}
  A.~Pilaftsis,
  Int.\ J.\ Mod.\ Phys.\ A {\bf 14}, 1811 (1999)
  [arXiv:hep-ph/9812256].

\bibitem{Spergel:2003cb}
D.~N.~Spergel {\it et al.}  [WMAP Collaboration],
Astrophys.\ J.\ Suppl.\  {\bf 148}, 175 (2003)
[arXiv:astro-ph/0302209].

\bibitem{Kolb:1990vq}
  E.~W.~Kolb and M.~S.~Turner,
Redwood City, USA: Addison-Wesley (1990) 547 p. (Frontiers in physics, 69).

\bibitem{Khlebnikov:1988sr}
  S.~Y.~Khlebnikov and M.~E.~Shaposhnikov,
  Nucl.\ Phys.\ B {\bf 308}, 885 (1988); J.~A.~Harvey and M.~S.~Turner,
  Phys.\ Rev.\ D {\bf 42}, 3344 (1990).

\bibitem{Nakaya:2002ki}
  T.~Nakaya  [SUPER-KAMIOKANDE Collaboration],
  eConf {\bf C020620}, SAAT01 (2002)
  [arXiv:hep-ex/0209036]; S.~Fukuda {\it et al.}  [Super-Kamiokande Collaboration],
  Phys.\ Lett.\ B {\bf 539}, 179 (2002)
  [arXiv:hep-ex/0205075];  Q.~R.~Ahmad {\it et al.}  [SNO Collaboration],
  Phys.\ Rev.\ Lett.\  {\bf 89}, 011302 (2002)
  [arXiv:nucl-ex/0204009]; K.~Eguchi {\it et al.}  [KamLAND Collaboration],
  Phys.\ Rev.\ Lett.\  {\bf 90}, 021802 (2003)
  [arXiv:hep-ex/0212021];  M.~Apollonio {\it et al.}  [CHOOZ Collaboration],
  Phys.\ Lett.\ B {\bf 466}, 415 (1999)
  [arXiv:hep-ex/9907037]; G.~L.~Fogli, E.~Lisi, A.~Marrone, A.~Palazzo and A.~M.~Rotunno,
      [arXiv:hep-ph/0506307].

\bibitem{pmns}
  B.~Pontecorvo,
  Sov.\ Phys.\ JETP {\bf 6}, 429 (1957), 
  Sov.\ Phys.\ JETP {\bf 7}, 172 (1958); 
  Z.~Maki, M.~Nakagawa and S.~Sakata,
  Prog.\ Theor.\ Phys.\  {\bf 28}, 870 (1962).




\bibitem{Ma:1998dx}
E.~Ma and U.~Sarkar,
Phys.\ Rev.\ Lett.\  {\bf 80}, 5716 (1998)
[arXiv:hep-ph/9802445]; G.~Lazarides and Q.~Shafi,
Phys.\ Rev.\ D {\bf 58}, 071702 (1998)
[arXiv:hep-ph/9803397]; T.~Hambye, E.~Ma and U.~Sarkar,
  Nucl.\ Phys.\ B {\bf 602}, 23 (2001)
  [arXiv:hep-ph/0011192].


\bibitem{Das:2000uk}
C.~R.~Das and M.~K.~Parida,
Eur.\ Phys.\ J.\ C {\bf 20}, 121 (2001)
[arXiv:hep-ph/0010004].

\bibitem{Klapdor-Kleingrothaus:2000sn}
H.~V.~Klapdor-Kleingrothaus {\it et al.},
Eur.\ Phys.\ J.\ A {\bf 12}, 147 (2001)
[arXiv:hep-ph/0103062];C.~E.~Aalseth {\it et al.}  [IGEX Collaboration],
  Phys.\ Rev.\ D {\bf 65}, 092007 (2002)
  [arXiv:hep-ex/0202026].

\bibitem{Haba:1998fb}
N.~Haba, N.~Okamura and M.~Sugiura,
Prog.\ Theor.\ Phys.\  {\bf 103}, 367 (2000)
[arXiv:hep-ph/9810471]; J.~A.~Casas, J.~R.~Espinosa, A.~Ibarra and I.~Navarro,
Nucl.\ Phys.\ B {\bf 569}, 82 (2000)
[arXiv:hep-ph/9905381]; S.~Antusch, J.~Kersten, M.~Lindner and M.~Ratz,
Phys.\ Lett.\ B {\bf 538}, 87 (2002)
[arXiv:hep-ph/0203233].

\bibitem{Chacko:1997cm}
Z.~Chacko and R.~N.~Mohapatra,
Phys.\ Rev.\ D {\bf 58}, 015003 (1998)
[arXiv:hep-ph/9712359]; B.~Dutta and R.~N.~Mohapatra,
Phys.\ Rev.\ D {\bf 59}, 015018 (1999)
[arXiv:hep-ph/9804277]; B.~Dutta, R.~N.~Mohapatra and D.~J.~Muller,
Phys.\ Rev.\ D {\bf 60}, 095005 (1999)
[arXiv:hep-ph/9810443]; A.~Melfo and G.~Senjanovic,
Phys.\ Rev.\ D {\bf 68}, 035013 (2003)
[arXiv:hep-ph/0302216];
P.~Achard {\it et al.}  [L3 Collaboration],
Phys.\ Lett.\ B {\bf 576}, 18 (2003)
[arXiv:hep-ex/0309076].


\bibitem{Azuelos:2005uc}
K.~Huitu, J.~Maalampi, A.~Pietila and M.~Raidal,
Nucl.\ Phys.\ B {\bf 487}, 27 (1997)
[arXiv:hep-ph/9606311]; G.~Azuelos, K.~Benslama and J.~Ferland,
[arXiv:hep-ph/0503096].

\bibitem{Gursey:1975ki}
F.~Gursey, P.~Ramond and P.~Sikivie,
Phys.\ Lett.\ B {\bf 60}, 177 (1976); Y.~Achiman and B.~Stech,
Phys.\ Lett.\ B {\bf 77}, 389 (1978); Q.~Shafi,
Phys.\ Lett.\ B {\bf 79}, 301 (1978); G.~R.~Dvali and Q.~Shafi, 
Phys.\ Lett.\ B {\bf 326}, 258 (1994)
[arXiv:hep-ph/9401337]; G.~R.~Dvali and Q.~Shafi,
Phys.\ Lett.\ B {\bf 339}, 241 (1994)
[arXiv:hep-ph/9404334]; J.~E.~Kim,
Phys.\ Lett.\ B {\bf 564}, 35 (2003)
[arXiv:hep-th/0301177].
      

\end{thebibliography}
\end{document}